\documentclass[%
 aip,
 pop,%
 amsmath,amssymb,
 reprint,%
 color,
]{revtex4-1}
\usepackage{graphicx,color,dcolumn,bm}

\draft 

\begin{document}


\title{Gyrokinetic Large Eddy Simulations}

\author{P. Morel}
\author{A. Ba\~n\'on Navarro}
\author{M. Albrecht-Marc}
\author{D. Carati}

\affiliation{Statistical and Plasma Physics Laboratory, Universit\'e Libre de Bruxelles, Belgium}

\author{F. Merz}
\author{T. G\"orler}
\author{F. Jenko}
\affiliation{Max-Planck-Institut f\"ur Plasmaphysik, Boltzmannstr. 2, D-85748 Garching, Germany}

\date{\today}

\begin{abstract}
The Large Eddy Simulation (LES) approach is adapted to the study of plasma microturbulence in a fully
three-dimensional gyrokinetic system. Ion temperature gradient driven turbulence is studied with the
{\sc GENE} code for both a standard resolution and a reduced resolution with a model for the sub-grid
scale turbulence. A simple dissipative model for representing the effect of the sub-grid scales on the
resolved scales is proposed and tested. Once calibrated, the model appears to be able to reproduce most
of the features of the free energy spectra for various values of the ion temperature gradient.
\end{abstract}

\pacs{}

\maketitle 

\section{Introduction}

Turbulence in plasmas shares several general features with fluid turbulence as modeled by the
Navier-Stokes equation. In particular, microturbulence in a background magnetic field as described
by the gyrokinetic formalism is thought to be characterized by a forward perpendicular cascade of
free energy similar to the direct cascade of kinetic energy in the Richardson-Kolmogorov picture
of fluid turbulence~\cite{schekochihin_PPCF_2008, tatsuno_PRL_2009, plunk_thesis_2009, plunk_JFM_2010}. Although in plasma microturbulence, there does
not exist an inertial range in the strict sense of the word, recent gyrokinetic simulations show
that an asymptotically free, self-similar, and highly local cascade develops at high perpendicular
wavenumbers~\cite{banon_PRL_2011}. While dissipative processes (due to a coupling to damped eigenmodes~\cite{hatch_2011}) are also active at low wavenumbers, a significant fraction
of the free energy is transported to small spatial scales and dissipated there. In a simulation of
the complete turbulent cascade process, it is thus important, in principle, to capture all scales from
the energy injection range down to the smallest relevant dissipative scales. In certain situations,
such direct numerical simulations (DNS) can be computationally expensive or even unfeasible, however.
These DNS limitations have prompted the development of hybrid approaches mixing {\em ab initio}
computation and modeling. In particular, Large Eddy Simulation (LES) techniques have been devised
for simulating turbulent fluids at high Reynolds number \cite{smagorinsky_MWR_1963, germano_PoF_1991}. In these simulations, the large scales are
computed explicitly while the influence of the smallest scales is modeled. The aim of the present work
is to extend this technique to the gyrokinetic equations which describe microturbulence in magnetically
confined plasmas~\cite{brizard_RMP_2007,garbet_NF_2010}. 

As far as fluids are concerned, the main idea behind the development of LES techniques is the assumed existence of a universal regime for the small scales. Indeed, the large scales in a turbulent fluid are very much influenced by the geometry of the flow. It is thus {\em a priori} not easy, and probably impossible, to design a general model for these large scales. On the contrary, if the Reynolds number is large enough, the smallest scales are supposed to be independent of the geometry and should only be affected by the physical properties of the fluid. Hence, there is a reasonable hope that the small scales can be represented by a general model. In practice, however, the Reynolds number is not always sufficiently large to reach such a regime and the geometry of wall-bounded flows has to be taken into account in most LES studies of turbulent fluids. Nevertheless, for almost half a century, LES have proven their ability to significantly decrease the numerical effort required to reproduce the main feature of large scale turbulent flows~\cite{smagorinsky_MWR_1963, germano_PoF_1991}. More recently, such methods have also been applied successfully to turbulence in conducting fluids~\cite{agullo_PoP_2001, knaepen_PoF_2004}. 

In the area of gyrokinetic turbulence, LES techniques have been explored, for instance, by Smith and
Hammett~\cite{smith_PoP_1997}, considering hyperviscosity models for two-dimensional drift-wave turbulence
-- and the aim of the present study is to extend the LES methodology to gyrokinetics in three spatial
dimensions. In this context, it should be noted that plasma microturbulence is different from ordinary
fluid turbulence in that it may be driven by various mechanisms such as, e.g., the presence of an ion
temperature gradient (ITG) or an electron temperature gradient (ETG). It is thus to be expected that the
modeling has to be adapted to the drive mechanism (rather than to the geometry of the system). In the
present work, we will focus on the case of ITG turbulence.

The paper is organized as follows. The fundamental equations are discussed in Section~\ref{gyro}, and the
LES approach for gyrokinetics is presented in Section~\ref{LES} together with a simple dissipation model.
The calibration of the model is discussed for Cyclone Base Case parameters which is a standard ITG turbulence test case. An estimate of the truncated scales is proposed for various quantities in Section~\ref{estimate}. The robustness of the model in terms of parameter changes is analyzed in Section~\ref{robustness}, followed by a summarizing discussion in Section~\ref{discussion}.

\section{Gyrokinetic model \label{gyro}}

The LES approach can, of course, be studied in the context of a general gyrokinetic system, including
multiple particle species, electromagnetic fluctuations, collisions, general tokamak geometry, profile
variations, and the like, as it is generally used in
{\sc GENE}~\cite{jenko_PoP_2000,dannert_PoP_2005,merz_thesis_2009}. However, in order to simplify the
following discussions, we restrict to a reduced system here, working
with only one ion species and treating the electrons as adiabatic. At the same time, a simple
$\hat s$-$\alpha$ model geometry is employed, and collisional effects are neglected. Moreover, the
radially local version of {\sc GENE} is used which solves the gyrokinetic equations in a flux-tube
geometry~\cite{beer_PoP_1995, lapillonne_PoP_2009}, employing the field-aligned coordinates
$(x,y,z,v_\parallel,\mu)$. The derivation of the corresponding non-dimensional equations can be found in previous studies~\cite{merz_thesis_2009}. The resulting expressions read as follows. The time evolution
of the ion distribution function $f_{ki}$ in $k$ space is given by
\begin{equation}
\label{eq:Vlasov}
  \partial_t f_{ki} = L [f_{ki}] + N[f_{ki},f_{ki}] + D[f_{ki}]\,,
\end{equation}
where the linear term can be split into three contributions, $L[f_{ki}] = L_G[f_{ki}] + L_C[f_{ki}] +
L_\parallel[f_{ki}]$, with
\begin{eqnarray}
\label{def:linear-terms}
L_G[f_{ki}] & = & - \left ( \omega_{ni} + \left ( v_\parallel^2 + \mu B_0 - 3/2 \right ) \omega_{Ti} \right )
F_0 i k_y (J_0 \phi_k) \nonumber \\
L_C[f_{ki}] & = & - \frac{T_{0i} ( 2 v_\parallel^2 + \mu B_0 )}{Z_i T_{0e} B_0}
( K_x i k_x h_{ki} + K_y i k_y h_{ki} ) \nonumber \\
L_\parallel [f_{ki}] & = & - \frac{v_{Ti}}{2} \left ( \partial_z \ln{F_0} \, \partial_{v_\parallel} h_{ki}
- \partial_{v_\parallel} \ln{F_0} \, \partial_z h_{ki} \right ) \,.
\nonumber
\end{eqnarray}
Here, $h_{ki}$ is defined as the nonadiabatic part of the distribution function, $h_{ki} = f_{ki} +
Z_i F_{0i} \phi_k T_{0e} / T_{0i}$ where $Z_i$ denotes the charge number, $F_{0i}$ the background distribution 
function, $\phi$ the electrostatic potential and $T_{0e}, T_{0i}$ the electron and ion temperatures.
The first linear term $L_G$ represents the influence of the fixed ion
density ($\omega_{ni}$) and temperature ($\omega_{Ti}$) gradients, the second linear term $L_C$ describes
effects due to magnetic curvature, and the third linear term $L_\parallel$ contains the parallel dynamics
involving magnetic trapping as well as the linear Landau damping. Meanwhile, the nonlinear term $N$
represents the effect of the self-consistent electric field in the $\vec{E}\times\vec{B}$ drift of
charged particles,
\begin{align}
N[f_{ki},f_{ki}] = \sum_{k'}  \left ( k_x' k_y - k_x k_y' \right ) J_0 \phi_{k'} f_{(k - k') i}\,,
\nonumber 
\end{align}
and the dissipation term $D[f_{ki}]$ is given by
\begin{align}
D[f_{ki}] = - \left(a_x\, (i\, k_x)^n+a_y\, (i\, k_y)^n+a_z\, \partial_z^n+a_{v_\parallel}\, \partial_{v_\parallel}^n\right)f_{ki} \,,
\nonumber
\end{align} 
where typically $n=4$ is used, and the coefficients $a_x$, $a_y$, $a_z$, and $a_{v_\parallel}$ can be adapted
to each specific class of physical problems. In the local version of {\sc GENE} used here, unknowns are
Fourier transformed in the radial and poloidal directions, so that $x$ and $y$ are replaced, respectively,
by $k_x$ and $k_y$. The subscript `$k$' has been added to label such Fourier space quantities. Due to the
imposed quasi-neutrality, the electrostatic potential $\phi$ and the distribution function are related via
the linear equation
\begin{align}
\label{eq:Poisson}
Z_i^2 n_{i0} \frac{T_{0e}}{T_{0i}} \left ( 1 - \Gamma_0(b_i) \right ) \phi_k
+ n_{e0} \left ( \phi_k - \left < \phi_k \right >_{FS} \right ) &=\nonumber\\
&\hspace{-40truemm} Z_i n_{i0} \pi B_0 \int\, d\mu\, dv_\parallel\, J_0(\lambda) f_{ki}\,,
\end{align}
with $\lambda^2=2\, k_\perp^2\, \mu/B_0$ and $b_i =v_{Ti}^2 k_\perp^2/(2 \Omega_{ci}^2)$. The functions $J_0$
and $\Gamma_0(b_i)=\exp(-b_i)\,I_0(b_i)$ are, respectively, the Bessel and the scaled modified Bessel
functions of order zero, where $\Omega_{ci}$ is the ion cyclotron pulsation and $q_i = Z_i e$ their charge.
In the flux-tube geometry (symbolically defined by the metric coefficients~\cite{lapillonne_PoP_2009}
$g^{xx}$, $g^{xy}$ and $g^{yy}$), the amplitude of the perpendicular wave vector $k_\perp$ is given by
$k_\perp^2 =g^{xx} k_x^2 + 2 g^{xy} k_x k_y + g^{yy} k_y^2$ and depends on $z$ through the metric coefficients. 
Note that $\langle\phi\rangle_{FS}$ represents the flux surface average of the electric potential.

One property of the gyrokinetic equations which is of particular interest here is the conservation of the free energy by the nonlinear term\cite{watanabe_NF_2006,candy_PoP_2006,schekochihin_PPCF_2008}. The latter quantity is defined as
\begin{align}
\label{eq:fe-definition}
\mathcal{E}= n_{0i} \frac{T_{0i}}{T_{0e}} \int d\Lambda_k \frac{h_{-ki}\, f_{ki}}{2 F_{0i}} \, ,
\end{align} 
where $h_{-ki} = h_i (-k_x,-k_y,z,v_\parallel,t)$. The integration over the phase space of a given quadratic unknown $|X_k|^2 = X(k_x, k_y, z, v_\parallel, \mu, t) X(-k_x,-k_y,z,v_\parallel,\mu,t)$ is given by
\begin{align}
\int\, d\Lambda_k\, |X_k|^2 = \frac{1}{V} \sum_{k_x^{\hbox{\tiny{\sc DNS}}}} \sum_{k_y^{\hbox{\tiny{\sc DNS}}}} \int\, \pi\, dz\, dv_\parallel\, d\mu \, |X_k|^2 \, ,
\end{align} 
where the sum over $k_x^{\hbox{\tiny{\sc DNS}}}$ has to be understood as a sum from $k_x= (-N_x/2+1)\, \Delta k_x$ to $k_x=N_x/2\, \Delta k_x$ and the sum over $k_y^{\hbox{\tiny{\sc DNS}}}$ corresponds to a sum from $k_y=(- N_y/2+1 )\, \Delta k_y$ to $k_y=N_y/2\, \Delta k_y$. Here, $\Delta k_x = 2 \pi / L_x$ and $\Delta k_y = 2 \pi / L_y$ are the smallest wave vectors that can be used  to represent periodic functions in rectangular domain of size $L_x \times L_y$. In practice, due to the symmetry of Fourier transform, negative $k_y$ modes are given by complex conjugation of positive $k_y$ modes. The volume $V$ is defined in the chosen magnetic $s-\alpha$ equilibrium by
\begin{align}
V = \sum_{k_x^{\hbox{\tiny{\sc DNS}}}} \sum_{k_y^{\hbox{\tiny{\sc DNS}}}} \int dz / B_0 \, .
\end{align} 
The two-dimensional spectral density of the free energy is defined by
\begin{align}
\mathcal{E}^{k_x,k_y}&= \frac{n_{0i}\, T_{0i}}{V\,T_{0e}}  \int \pi dz dv_\parallel d\mu \, \biggl(\frac{h_{-ki}\, f_{ki}}{2 F_{0i}}\biggr) \, ,
\end{align} 
and the one-dimension spectral densities along $k_x$ or $k_y$ are simply given by
\begin{align}
\mathcal{E}^{k_x}=\sum_{k_y^{\hbox{\tiny{\sc DNS}}}} \mathcal{E}^{k_x,k_y}\,,
\hspace{10truemm}\mathcal{E}^{k_y}=\sum_{k_x^{\hbox{\tiny{\sc DNS}}}} \mathcal{E}^{k_x,k_y}\,.
\nonumber
\end{align} 
The free energy balance can be expressed as follows
\begin{equation}
\label{eq:fe-balance}
\partial_t \mathcal{E} = \mathcal{G} - \mathcal{D} \, .
\end{equation}
The two terms in the right hand side represent the free energy injection $\mathcal{G}$ and dissipation $\mathcal{D}$. They are given by
\begin{eqnarray}
\label{def:G}
\mathcal{G} & = & n_{0i} \frac{T_{0i}}{T_{0e}} \int  d\Lambda_k \frac{h_{-ki}}{F_{0i}} L_G [f_{ki}] \, , \\
\label{def:D}
\mathcal{D} & = & - n_{0i} \frac{T_{0i}}{T_{0e}} \int d\Lambda_k \frac{h_{-ki}}{F_{0i}} D [f_{ki}]\,.
\end{eqnarray} 
The free energy injection term $\mathcal{G}$ is directly related to the ion heat flux, $\mathcal{Q}_i$:
\begin{align}
\mathcal{G} = \omega_{Ti} \mathcal{Q}_i \, .
\label{eq:heat-flux}
\end{align}
The heat diffusivity $\chi_i$ and heat flux $\mathcal{Q}_i = n_{0i} T_{0i} \omega_{Ti} \chi_i$ are considered as reference quantities for comparison between gyrokinetic numerical solvers as well as with experiments. The appearance of heat flux as the free energy source stresses the importance of free energy balance in gyrokinetics.

\section{\label{LES} LES for gyrokinetics}

\begin{figure}
\includegraphics[width=0.5\textwidth]{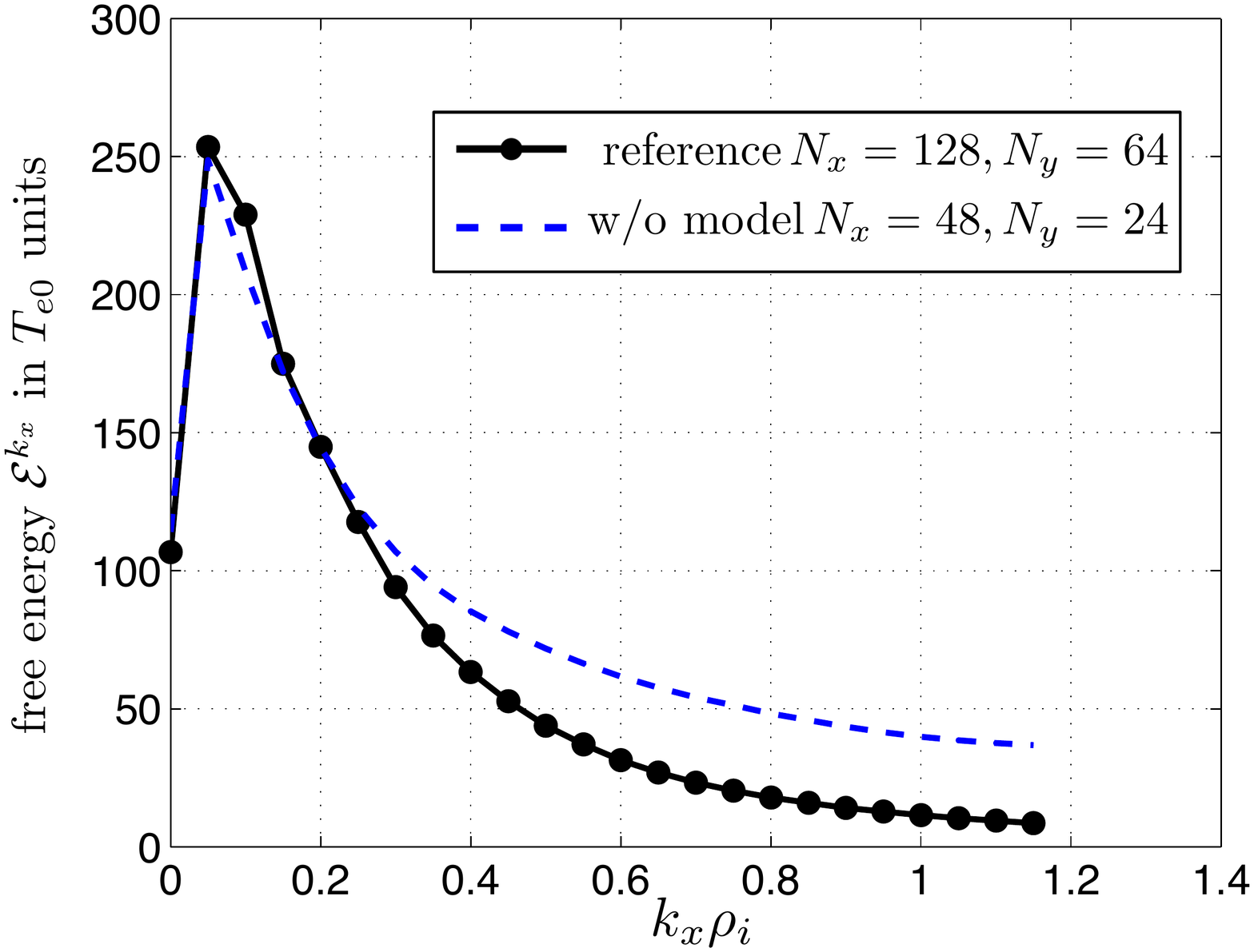}
\caption{\label{fig:cmp-FE-CBC} Free energy: comparison between highly resolved DNS (black), and LES without model (blue) for the cyclone base test case. The resolution for the DNS is $N_x=128$ and $N_y=64$ and for the LES $N_x=48$ and $N_y = 24$.}
\end{figure}

The main objective of the LES technique is to explore the large scale physics at a lower computational
cost when compared to DNS. Reducing the cost of a gyrokinetic simulation can be achieved by several ways.
In a Eulerian approach, the distribution function is represented on a fixed grid in five-dimensional phase
space, using $N_x\times N_y\times N_z\times N_{v_\parallel}\times N_\mu$ grid points. In a gyrokinetic LES, this
grid is then to be coarsened. Considering that the perpendicular cascade processes are expected to transfer
free energy from large spatial scales to small ones, we just employ the technique to the $(x,y)$ grid in the
present study. In the following, the DNS and LES grids correspond, respectively, to $N_x\times N_y=
128\times 64$ and $\overline{N}_x\times\overline{N}_y=48\times 24$, while $N_z=16$, $N_{v_\parallel}=32$, and
$N_\mu=8$ are held constant. The coarsening of the grid can be viewed as a low pass filter, denoted hereafter
by the $\overline{\cdots}$ symbol, that sets to zero the highest $k_x$ and $k_y$ modes. The filtered
distribution function will thus be labeled $\overline{f}_{ki}$. Applying this filter to the gyrokinetic
equation yields:
\begin{align}
\label{eq:filter-Vlasov}
\partial_t \overline{f}_{ki}  =  L[\overline{f}_{ki}] + N[\overline{f}_{ki}, \overline{f}_{ki}]
+ D[\overline{f}_{ki}] + \overline{T} \, ,
\end{align}
which contains a term $\overline{T}$ that depends explicitly on both the filtered distribution $\overline{f}_{ki}$ and on the unfiltered distribution $f_{ki}$
\begin{align}
\label{eq:subgridterm}
\overline{T} = \overline{N[f_{ki}, f_{ki}]} - N[\overline{f}_{ki}, \overline{f}_{ki}]\,.
\end{align} 
Except for the presence of $\overline{T}$ on the right hand side, Eq.~(\ref{eq:filter-Vlasov}) for
$\overline{f}_{ki}$ has the same form as Eq.~(\ref{eq:Vlasov}) for $f_{ki}$. The term $\overline{T}$
is usually referred to as the sub-grid scale term, though in the present situation, the terminology
sub-filter scale term would be more appropriate. In order to close Eq.~(\ref{eq:filter-Vlasov}),
$\overline{T}$ must be approximated by a model that has to be expressed in terms of the filtered
distribution $\overline{f}_{ki}$:
\begin{align}
\label{eq:subgridterm}
\overline{T} \approx M[\overline{f}_{ki}]\,.
\end{align} 
The importance of the sub-grid scale term is illustrated in Fig. \ref{fig:cmp-FE-CBC}. Two free energy
spectra are represented for the Cyclone Base Case~\cite{dimits_PoP_2000} (CBC) for ITG driven turbulence
($\omega_{ni} = 2.22$, $\omega_{Ti} = 6.92$, $q = 1.4$, $\hat{s} = 0.796$, $\epsilon = 0.18$,
$T_{e0}/T_{i0}=1$, $Z_i=1$). The perpendicular box sizes are given by $L_x = L_y= 125\, \rho_i$. In {\sc GENE}, numerical dissipation~\cite{pueschel_CPC_2010} can be introduced via fourth-order
derivatives along $z$ and $v_\parallel$:
\begin{align}
D[f_{ki}] = - a_z \partial_z^4 f_{ki} - a_{v_\parallel} \partial_{v_\parallel}^4 f_{ki}\,,
\end{align}
where the values for the coefficients $a_z$ and $a_{v_\parallel}$ are to be adjusted appropriately. In
Fig.~\ref{fig:cmp-FE-CBC}, the DNS spectrum is compared to the LES spectrum obtained by setting
$\overline{T}=0$. Clearly, the free energy is piling up in the high $k_x$ range in the latter case
due to the reduced high-wavenumber dissipation in the absence of small scales. As a secondary effect,
the free energy appears to be pumped out more rapidly of the large scales where the gradient source
term is active. Indeed, the free energy is transferred to the small scales by the nonlinear term.
These transfers have been identified as mostly local in Fourier space~\cite{banon_PRL_2011}.
Consequently, the increase of activity of modes closer to the injection range can explain that the
free energy is removed from the drive range more rapidly. Such a scenario is reminiscent of what is
observed in underresolved DNS of Navier-Stokes turbulence, where the energy is also piling up in the
large wavenumber range.

\begin{figure}
\includegraphics[width=0.50\textwidth]{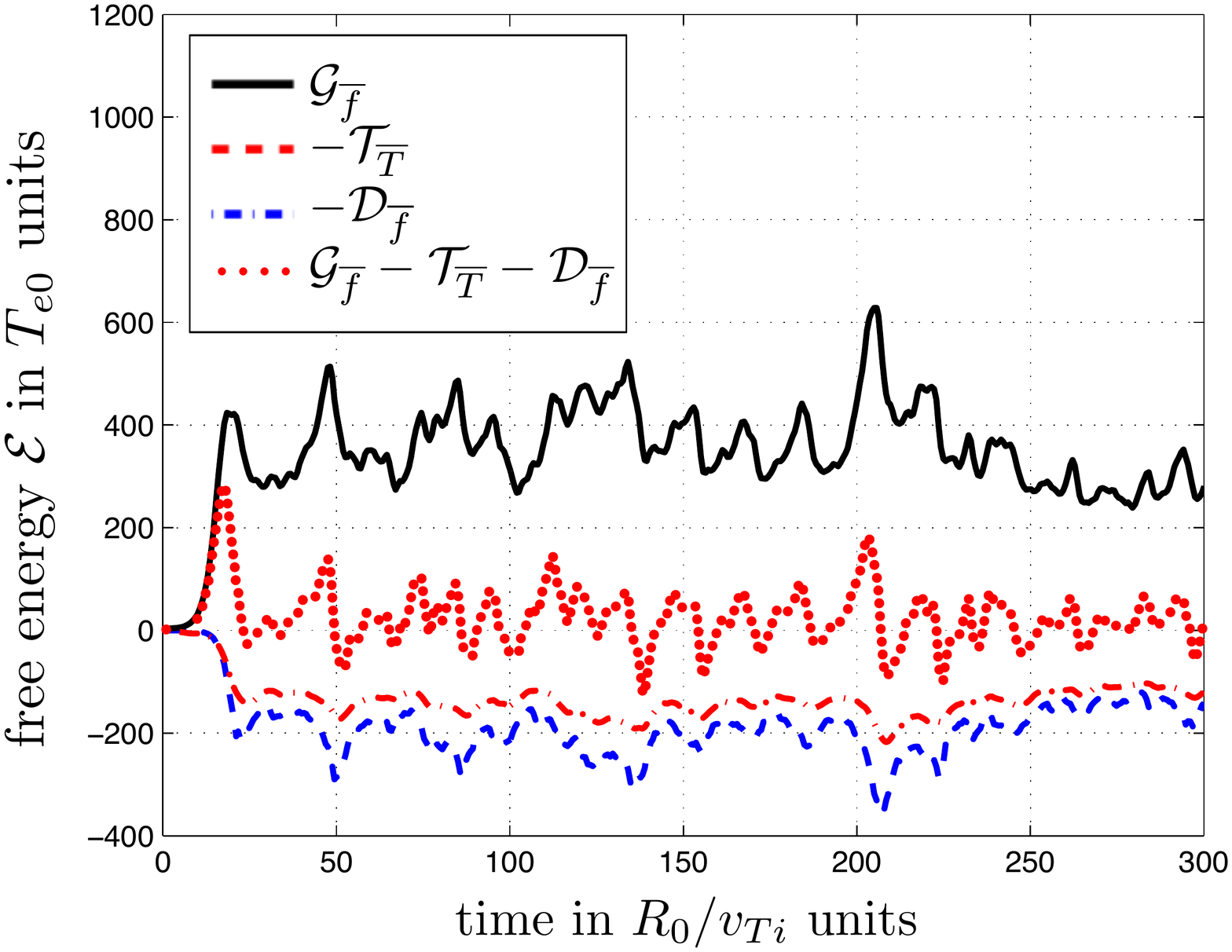}
\caption{\label{fig:sub-grid-FE} Sub grid contribution to the free energy balance compared with the free energy injection and dissipation terms. CBC parameters with a filter corresponding to $\overline{N}_x = 48$, $\overline{N}_y = 24$.}
\end{figure}

In Fig.~\ref{fig:cmp-FE-CBC}, the LES has been performed with $\overline{T}=0$, which can be considered
as the simplest sub-grid scale model. Obviously, such a choice is too simple since the free energy spectrum
deviates significantly from the DNS observations. The role of $\overline{T}$ can be understood by
considering the resolved free energy balance. Since $f_{ki}$ and $\overline{f}_{ki}$ satisfy the same
equation up to the term $\overline{T}$, the free energy associated to $\overline{f}_{ki}$, referred to as
the resolved free energy in the framework of a LES, must satisfy the following equation:
\begin{equation}
\label{eq:rfe-balance}
\partial_t \overline{\mathcal{E}}= \overline{\mathcal{G}} - \overline{\mathcal{D}} - \overline{\mathcal{T}}
\end{equation}
where the quantities $\overline{\mathcal{E}}$, $\overline{\mathcal{G}}$, $\overline{\mathcal{D}}$ are the same as $\mathcal{E}$, $\mathcal{G}$, $\mathcal{D}$, except that they are defined using $\overline{f}_{ki}$ and $\overline{h}_{ki}$ instead of $f_{ki}$ and $h_{ki}$. It should be noted, however, that all these global quantities are defined using a volume integration over $d\overline{\Lambda}_k$ in which the sums are over $k_x^{\hbox{\tiny{\sc LES}}}$ and $k_y^{\hbox{\tiny{\sc LES}}}$ and have to be understood as a sum from $k_x= ( -\overline{N}_x/2+1)\, \Delta k_x $ to $k_x= \overline{N}_x/2\, \Delta k_x$ and from $k_y=(-\overline{N}_y/2)\, \Delta k_y$ to $k_y= \overline{N}_y/2\, \Delta k_y$. Since the computational box sizes are the same in the LES and in the reference DNS, the same grid spacings $\Delta k_x$ and $\Delta k_y$ are used in both LES and DNS runs. However, the largest wave vectors are smaller in the LES than in the DNS: $K_x^{\hbox{\tiny{\sc LES}}}=\overline{N}_x/2\, \Delta k_x<K_x^{\hbox{\tiny{\sc DNS}}}=N_x/2\, \Delta k_x$ and 
$K_y^{\hbox{\tiny{\sc LES}}}=\overline{N}_y/2\, \Delta k_y<K_y^{\hbox{\tiny{\sc DNS}}}=N_y/2\, \Delta k_y$. The new term $\mathcal{T}_{\overline{T}}$ is defined by
\begin{equation}
\label{eq:rfe-balance}
\mathcal{T}_{\overline{T}} = -\int d\overline{\Lambda}_k \, n_{0i} \frac{T_{0i}}{T_{0e}} \frac{\overline{h}_{-ki}}{F_{0i}} \, \overline{T}
\end{equation}
and represents the effect of the sub-grid scales on the resolved free energy. If the cascade picture applies, the effect of the $\mathcal{T}_{\overline{T}}$ should be to pump out the resolved free energy in order to mimic the transfer towards the unresolved scales. If $f_{ki}$ and $h_{ki}$ are known from a DNS, the term $\overline{T}$ and consequently $\mathcal{T}_{\overline{T}}$ can be computed exactly. Using the same parameter as in Fig.~\ref{fig:cmp-FE-CBC}, $\mathcal{T}_{\overline{T}}$ has been computed and is shown in Fig.~\ref{fig:sub-grid-FE}. It is indeed negative and represents a loss of resolved free energy. Its amplitude is compared to the resolved free energy injection rate $\mathcal{G}_{\overline{f}}$ and dissipation rate $\mathcal{D}_{\overline{f}}$. On average, once turbulence is developed and a statistically stationary regime is reached, these three terms should be in balance, $\mathcal{G}_{\overline{f}}\approx\mathcal{D}_{\overline{f}}+\mathcal{T}_{\overline{T}}$. In the run corresponding to Fig.~\ref{fig:cmp-FE-CBC}, the ratio $\mathcal{T}_{\overline{T}}/\mathcal{D}_{\overline{f}}$ appears to be close to unity. Hence, the transfer of free energy between the resolved and the unresolved scales cannot be neglected. 

The development of models for representing the effect of small, under-resolved scales on the large, resolved scales has been the subject of countless efforts in LES for fluid turbulence. However, the most commonly used models simply attempt to reproduce the transfer of kinetic energy towards the unresolved scales by a dissipative mechanism usually represented by an effective viscosity. Considering the analogy between fluid and plasma turbulence, it is proposed here to also use an effective dissipation which is modeled by the hyper-diffusion term
\begin{align}
\label{def:model}
M[\overline{f}_{ki}] = - c_\perp k_\perp^4 \overline{h}_{ki}\,.
\end{align}
It is easy to verify that such a model always gives a negative contribution to the resolved free energy balance:
\begin{align}
\label{TM}
\mathcal{T}_M & = - n_{0i} \frac{T_{i0}}{T_{e0}}\int d\overline{\Lambda}_k \frac{\overline{h}_{-ki}}{F_{0i}} M[\overline{h}_{ki}] \nonumber \\
& \hspace{10truemm} = - c_\perp n_{0i} \frac{T_{i0}}{T_{e0}} \int d\overline{\Lambda}_k \left | \frac{k_\perp^2 \overline{h}_{ki}}{\sqrt{F_{0i}}} \right |^2 < 0\,.
\end{align}
The hyper-diffusion coefficient $c_\perp$ can be adjusted by comparing the results given by a reference (well resolved) DNS with results from an LES using Eq.~(\ref{TM}). 

\begin{figure}
\includegraphics[width=0.50\textwidth]{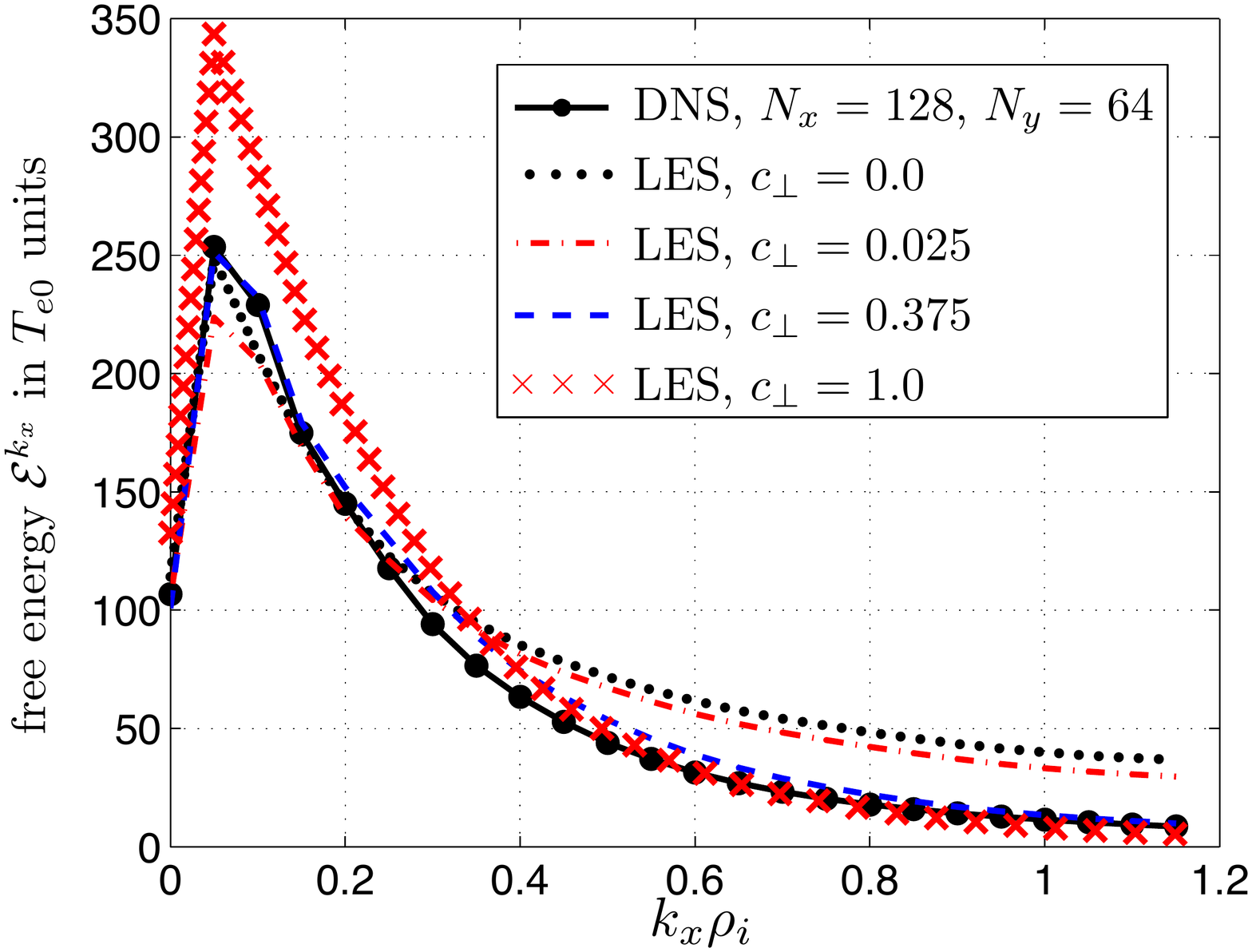}
\includegraphics[width=0.50\textwidth]{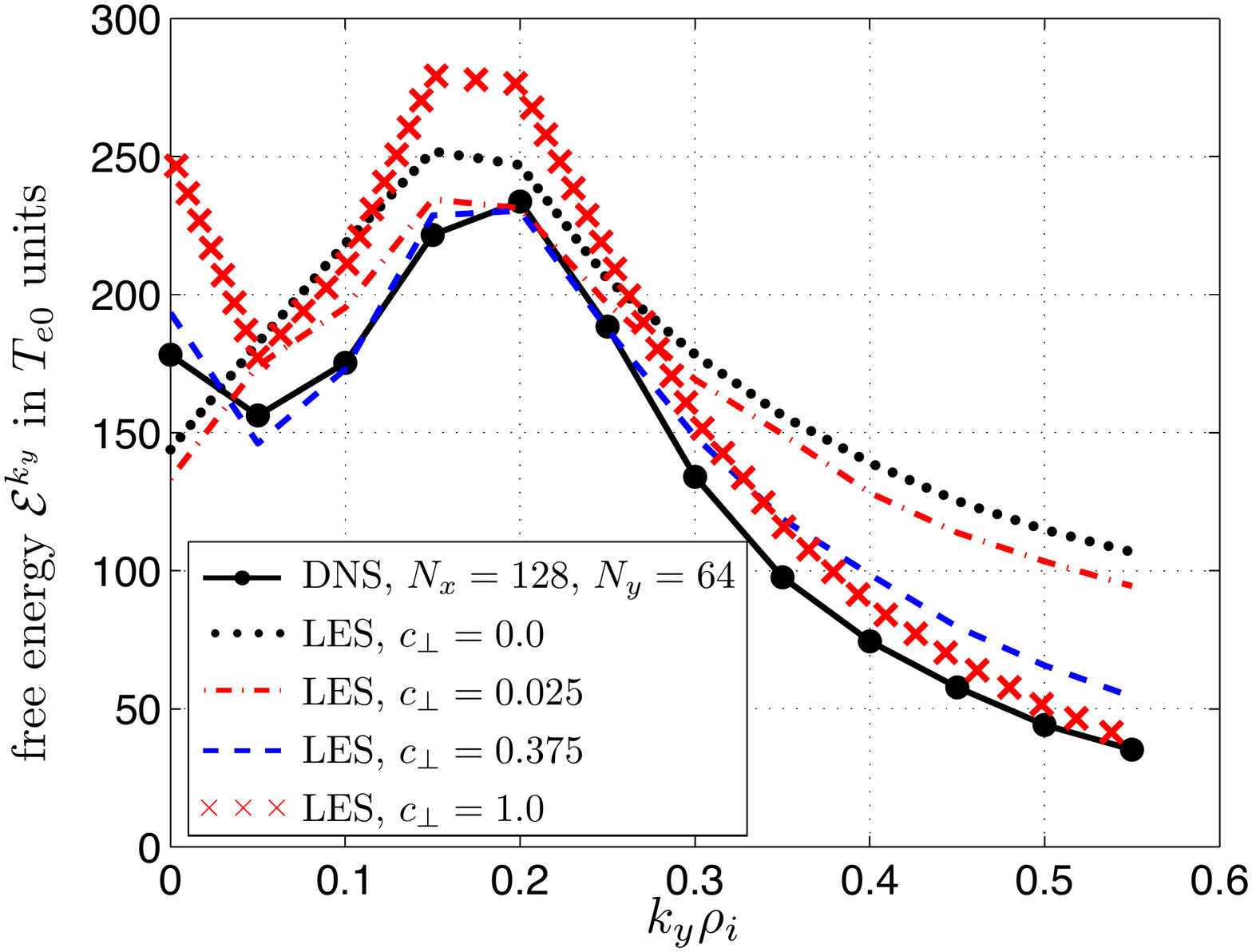}
\caption{\label{fig:scan-cperp} Resolved free energy spectra $\mathcal{E}^{k_x}$ (top) and $\mathcal{E}^{k_y}$ (bottom) obtained by varying the model coefficient compared with reference DNS.}
\end{figure}

In Fig.~\ref{fig:scan-cperp}, the free energy spectra $\mathcal{E}^{k_x}$ and $\mathcal{E}^{k_y}$, are displayed for various values of $c_\perp$. The black curve corresponds to the reference DNS run. It is compared to four LES runs corresponding to $c_\perp = \{0, 0.025, 0.375, 1.0 \}$. Other values of $c_\perp$ have also been tested but are not shown here for clarity. All the spectra are time-averaged during the turbulent phase over a period of $2000 R_0 / v_{Ti}$. Obviously, a too small parameter ($c_\perp=0.025$ dash-dotted) does not significantly improve the result when compared to the no-model case ($c_\perp=0$ dotted). Also, a too large value of $c_\perp$ ($\times$ dotted) tends to over-damp the small scales which leads to an artificial accumulation of free energy in the large scales (small $k$). The optimal value appears to be close to $c_\perp=0.375$ (dashed). The corresponding LES reproduces fairly well the spectra of the resolved free energy both in $k_x$ and in $k_y$.

\section{\label{estimate} Estimate for the sub-grid quantities}

The use of a model has been shown in the preceding section to improve significantly the agreement between DNS and LES in gyrokinetic simulations. However, since small scales are truncated in the LES runs, it is not possible in LES to predict directly global quantities such as the total free energy, the total heat flux (or equivalently, the total free energy injection) and the total free energy dissipation. An estimate of the contribution from the truncated scales to these global quantities is certainly desirable if a comparison has to be made with experimental results.

In this section, a simple estimate is proposed for the subgrid scale contribution to these quantities. It is noted that all these quantities (${\cal E}$, ${\cal G}$ or ${\cal D}$), generically represented by $Q$, can be represented either by the their two-dimensional spectrum $Q^{k_x,k_y}$ or by their one-dimensional $k_x$ spectrum ($Q^{k_x}$) and $k_y$ spectrum($Q^{k_y}$). In a LES, only the resolved part of $Q$, denoted hereafter $\overline{Q}$ is directly accessible. It is given by:
\begin{align}
\overline{Q}=\sum_{|k_x|\leq K_x^{\hbox{\tiny{\sc LES}}}} Q^{k_x}=\sum_{|k_y|\leq K_y^{\hbox{\tiny{\sc LES}}}} Q^{k_y}\,.
\label{eq:Qbar}
\end{align}
The unresolved part of $Q$, denoted $\delta Q$, contains three contributions $\delta Q=\delta_x Q+\delta_y Q+\delta_{xy} Q$:
\begin{align}
\delta_x Q&=\sum_{|k_x|>K_x^{\hbox{\tiny{\sc LES}}}} \sum_{|k_y|\leq K_y^{\hbox{\tiny{\sc LES}}}}\, Q^{k_x,k_y}\,,\\
\delta_y Q&=\sum_{|k_x|\leq K_x^{\hbox{\tiny{\sc LES}}}} \sum_{|k_y|> K_y^{\hbox{\tiny{\sc LES}}}}\, Q^{k_x,k_y}\,,\\
\delta_{xy} Q&=\sum_{|k_x|>K_x^{\hbox{\tiny{\sc LES}}}} \sum_{|k_y|>K_y^{\hbox{\tiny{\sc LES}}}}\, Q^{k_x,k_y}\,.
\label{eq:Qdelta}
\end{align}
In DNS, $\delta Q$ can be computed, but, in LES, it has to be estimated. Such an estimate can  be obtained by noting that, in the large $k_x$ and $k_y$ ranges of LES runs, the quantity $Q$ can often be approximated by decaying power laws:
\begin{align}
Q^{k_x}\approx A_x\, k_x^{-\alpha_x} \, , \hspace{10truemm} Q^{k_y}\approx A_y\, k_y^{-\alpha_y} \, .
\label{eq:Qpower}
\end{align}
The amplitudes $A_x$ and $A_y$ as well as the exponents $\alpha_x$ and $\alpha_y$ can be estimated by linear regression from the LES spectra. In that case, the following estimates can be obtained~:
\begin{align}
\delta_x Q&\approx\sum_{|k_x|>K_x^{\hbox{\tiny{\sc LES}}}}^{K_x^{\hbox{\tiny{\sc DNS}}}}\,A_x\, k_x^{-\alpha_x}\label{eq:Qdeltaestx} \, ,
\\
\delta_y Q&\approx\sum_{|k_y|>K_y^{\hbox{\tiny{\sc LES}}}}^{K_y^{\hbox{\tiny{\sc DNS}}}}\,A_y\, k_y^{-\alpha_y} \, .
\label{eq:Qdeltaesty}
\end{align}
Since these sums are finite, there is a priori no restriction on the values of the exponents. However, if the wave vector range is extended to infinity, these sums converge if and only if $\alpha_x>1$ and $\alpha_y>1$. Estimating $\delta_{xy} Q$ is more difficult. However, assuming a separable spectrum $Q^{k_x,k_y}=q_1^{k_x}\,q_2^{k_y}$, it can be shown that $\delta_{xy} Q=\delta_x Q\ \delta_y Q/\overline{Q}$. In general, it is thus expected that the correction due to $\delta_{xy} Q$ is very small and can be neglected compared to $\delta_x Q$ or $\delta_y Q$. This procedure has been used to estimate the total value of both the free-energy and the heat flux from the LES with the optimal value of $c_\perp$ and for the LES without model.

\begin{figure}[ht]
\includegraphics[width=0.50\textwidth]{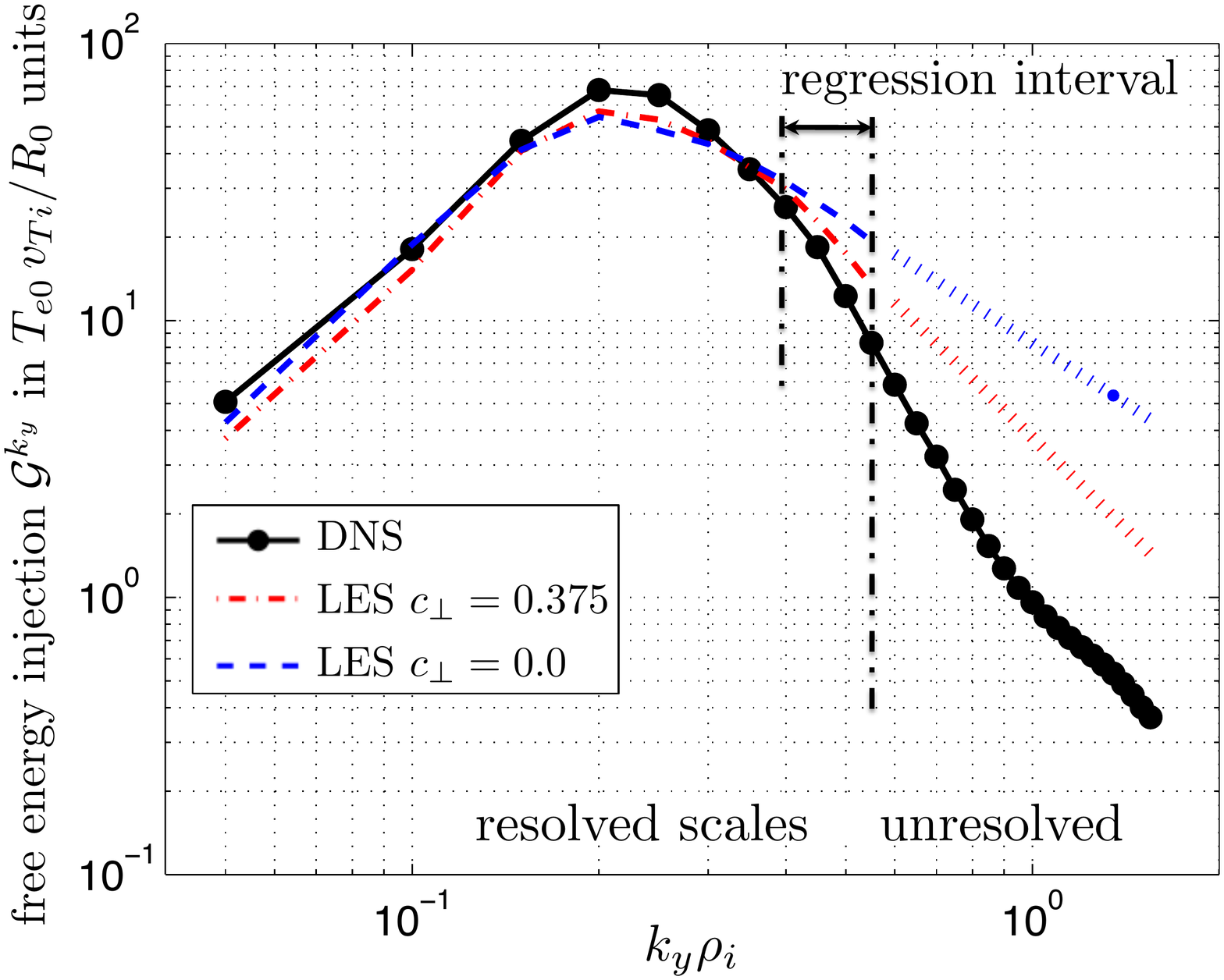}
\caption{\label{fig:spectral-injection} Free energy injection spectra $\mathcal{G}^{k_y}$ obtained with the model $c_\perp = 0.375$, compared with reference DNS and without model. Unresolved spectra of LES $c_\perp = 0.375$ and $c_\perp = 0.0$ have been constructed using  a regression method with the assumption (\ref{eq:Qpower}).}
\end{figure}

As far as the free energy is concerned, the best fit of the one-dimensional spectra yields $\alpha_x = 2.21$, $\alpha_y = 1.74$ for $c_\perp=0.375$. These values are then used to compute the corrections $\delta_x{\cal E}$ and $\delta_y{\cal E}$. The LES estimate ${\cal E}^{\hbox{\tiny{\sc LES}}}$ for the total free energy can then be compared to the value measured from the DNS, ${\cal E}^{\hbox{\tiny{\sc DNS}}}$. It is found that the LES estimate is in good agreement with the DNS value ${\cal E}^{\hbox{\tiny{\sc LES}}} = 1.10\, {\cal E}^{\hbox{\tiny{\sc DNS}}}$. On the contrary, without a model ($c_\perp=0$), the best fit of the free-energy one dimensional spectra yields $\alpha_x = 0.79$, $\alpha_y = 0.85$. In that case, the estimates~(\ref{eq:Qdeltaestx}) and~(\ref{eq:Qdeltaesty}) would be divergent if the sums had to be extended to infinity. However, if the sums are limited to $K^{\hbox{\tiny{\sc DNS}}}$, it is possible to reconstruct the total free energy from the LES without model but the estimate is more than twice the value of the DNS: ${\cal E}^{\hbox{\tiny{\sc No Model}}} = 2.1\, {\cal E}^{\hbox{\tiny{\sc DNS}}}$.

The same procedure has been used for the free energy injection $\mathcal{G} = \omega_{Ti}\, \mathcal{Q}_i$. For $c_\perp=0.375$, the regression method yields $\alpha_x= 3.60$ and $\alpha_y= 2.20$, which gives the following estimate ${\cal G}^{\hbox{\tiny{\sc LES}}} = 1.11\, {\cal G}^{\hbox{\tiny{\sc DNS}}}$. Hence, again the value computed from the LES slightly overestimate the DNS value of the free energy injection. However this prediction is still in reasonable agreement with the DNS and provides a much better estimate than the no-model simulation for which ${\cal G}^{\hbox{\tiny{\sc No Model}}} = 1.38\, {\cal G}^{\hbox{\tiny{\sc DNS}}}$.

\section{\label{robustness} Robustness of the LES approach}

The choice $c_\perp = 0.375$ has proven to give a reasonable agreement between the LES and the DNS predictions, in the case of standard CBC parameters. However, the LES methodology is only useful if the model parameters don't have to be calibrated for each set of parameters. In this section, it is proposed to explore the robustness of the LES approach by varying the logarithmic temperature gradient $\omega_{Ti}$.

\begin{figure}
\includegraphics[width=0.50\textwidth]{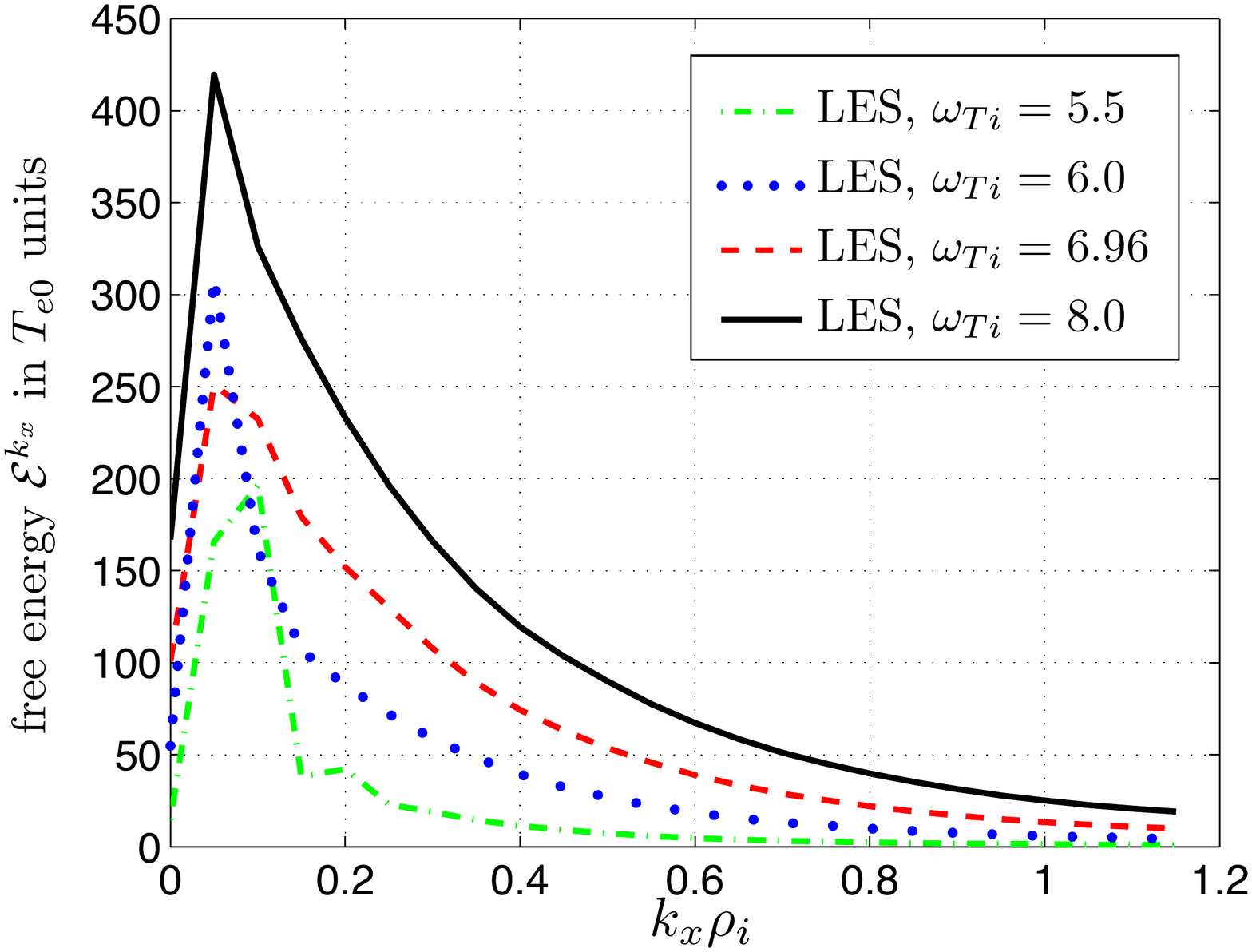}
\includegraphics[width=0.50\textwidth]{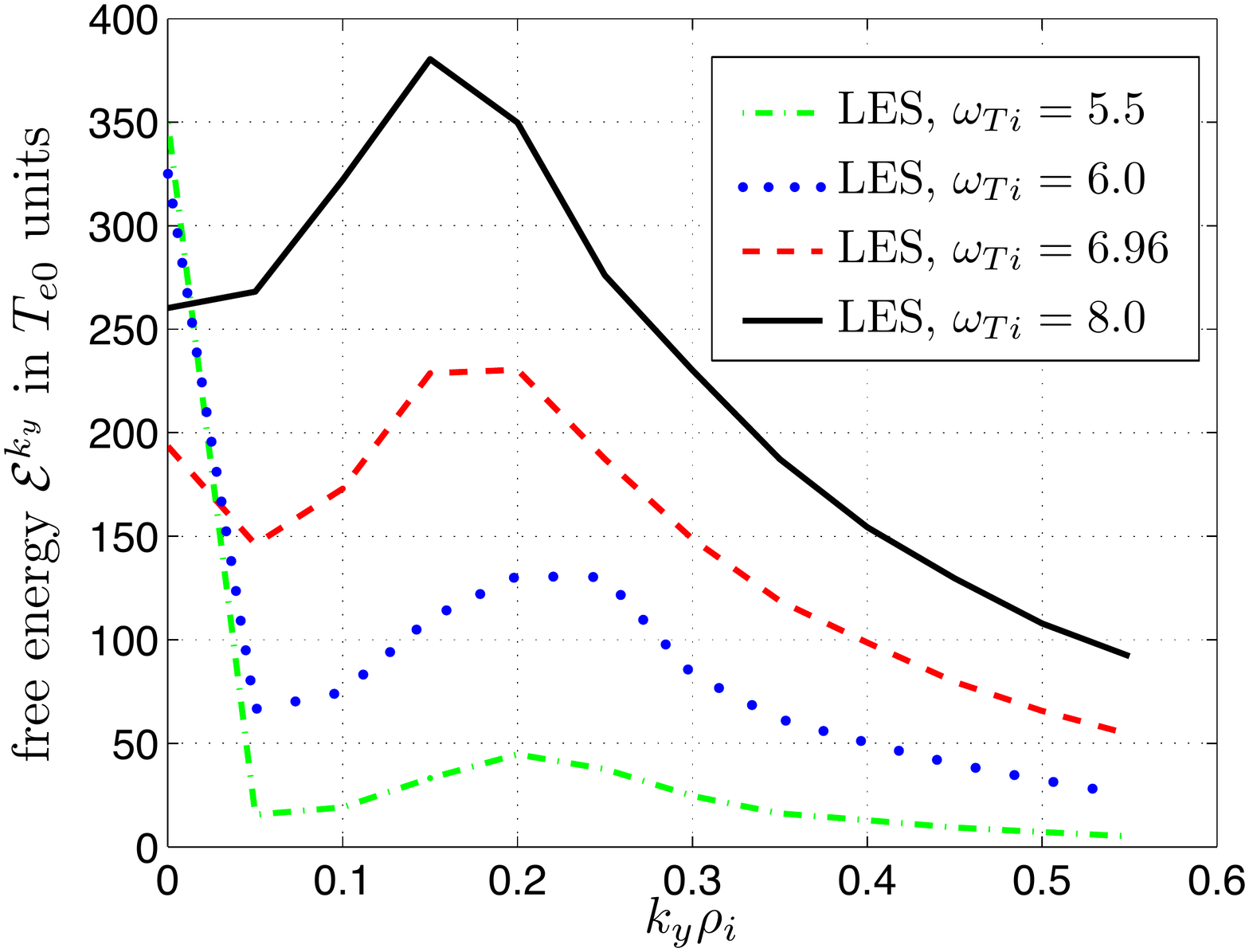}
\caption{\label{fig:scan-LT} Resolved free energy spectra $\mathcal{E}^{k_x}$ (top) and $\mathcal{E}^{k_y}$ (bottom)  obtained with the model~(\ref{def:model}) $c_\perp = 0.375$ for various values of the logarithmic temperature gradient $\omega_{Ti}$.}
\end{figure}

One of the most well known features of ITG turbulence is the Dimits shift~\cite{dimits_PoP_2000}, i.e., a nonlinear upshift of the stability threshold with respect to a linear analysis. This upshift occurs when varying the values of the logarithmic temperature gradient $\omega_{Ti}$, while keeping the logarithmic density gradient $\omega_{ni}$ constant. The explanation of such an effect is that turbulence nonlinearly transfers the free energy to the zonal flows (i.e., purely radial structures, corresponding to finite $k_x$, but
$k_y = k_\parallel = 0$). These structures then suppress the ITG instability if its linear growth rate is not sufficiently large, and turbulence can not be driven even if the plasma is linearly unstable.

In Fig.~\ref{fig:scan-LT}, all the parameters characterizing the CBC have been kept constant, except the logarithmic temperature gradient which is varied from the standard CBC value of $\omega_{Ti} = 6.96$  to $\omega_{Ti} = 5.5 ; 6.0 ; 8.0$. All simulations are performed using the model described in Eq.~(\ref{def:model}), with $c_\perp = 0.375$. 
The values $\omega_{Ti} = 6.0$ and $\omega_{Ti} = 5.5$ are close to the nonlinear threshold. It is then observed that an important part of the free energy is stored into the zonal flows. Such a result is in qualitative agreement with the usual picture of the Dimits shift~\cite{itoh_PoP_2006}. On the contrary when the temperature gradient is increased, the total free energy increases, and peaks around $k_x \rho_i \sim 0.05$, $k_y \rho_i \sim 0.15$.

\begin{figure}
\includegraphics[width=0.50\textwidth]{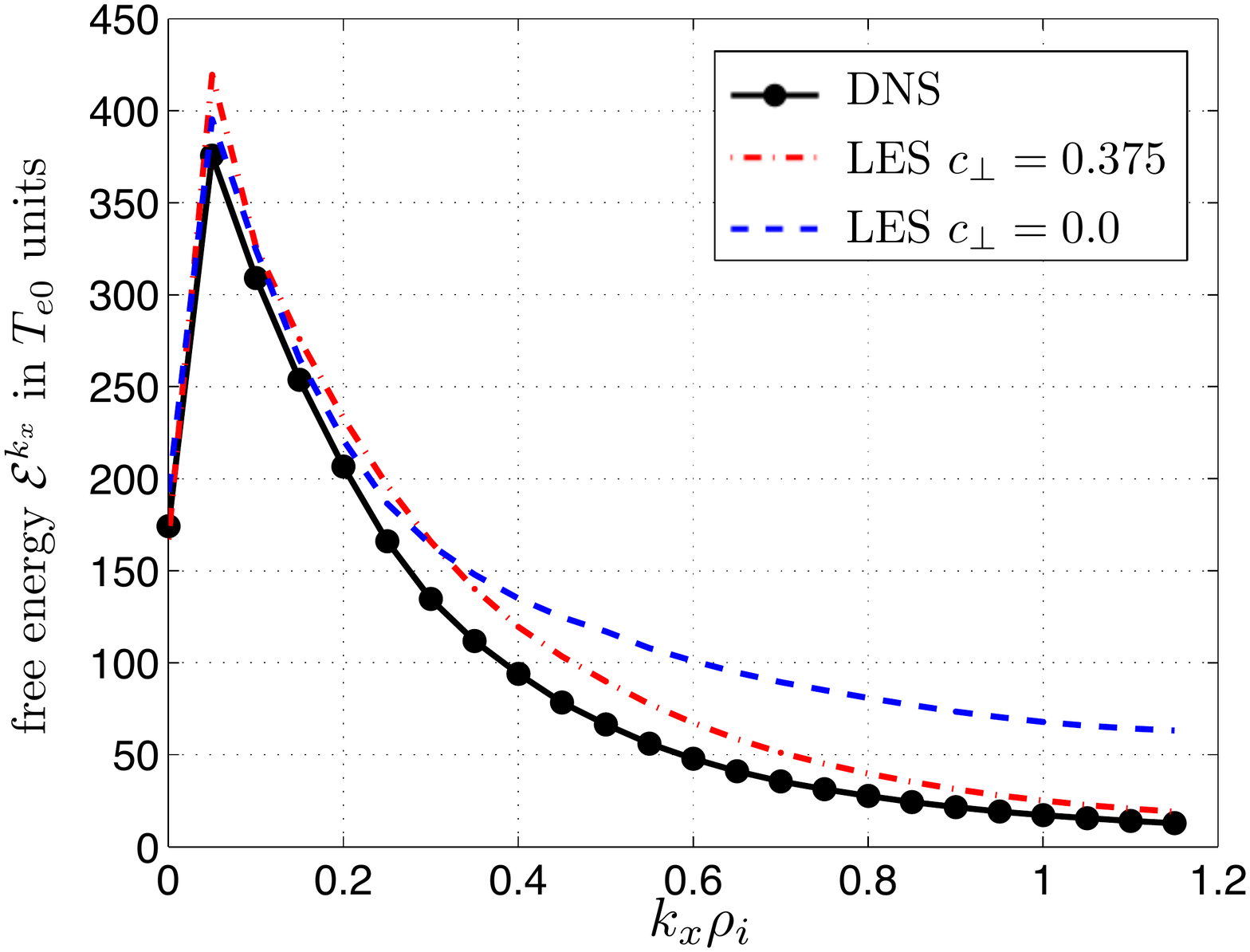}
\includegraphics[width=0.50\textwidth]{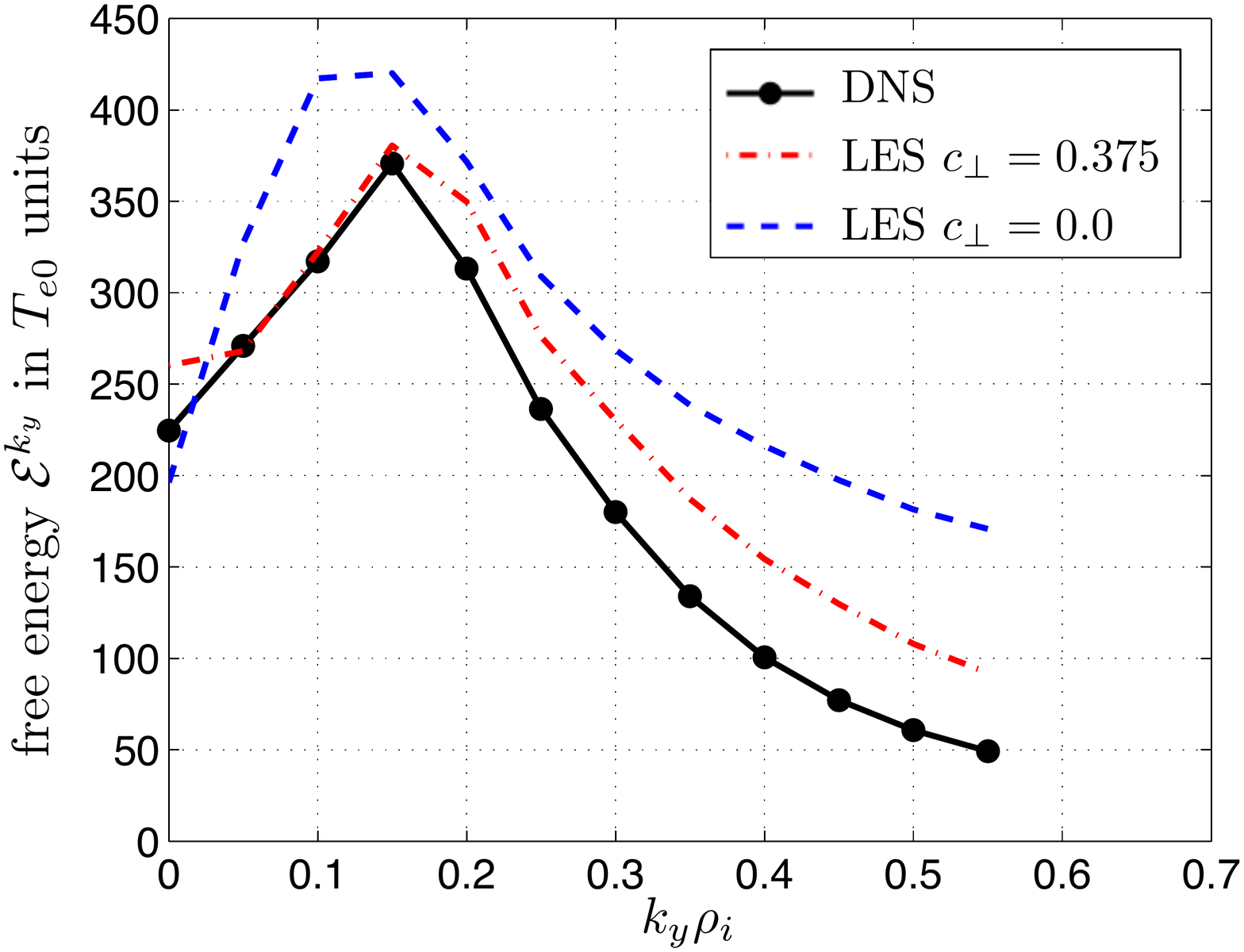}
\caption{\label{fig:cmp-LT80} Resolved free energy spectra $\mathcal{E}^{k_x}$ (top) and $\mathcal{E}^{k_y}$ (bottom)  obtained with the model~(\ref{def:model}) $c_\perp = 0.375$  for $\omega_{Ti} = 8.0$. Comparison with a highly resolved DNS with $\omega_{Ti} = 8.0$ and with LES without model.}
\end{figure}

\begin{figure}
\includegraphics[width=0.50\textwidth]{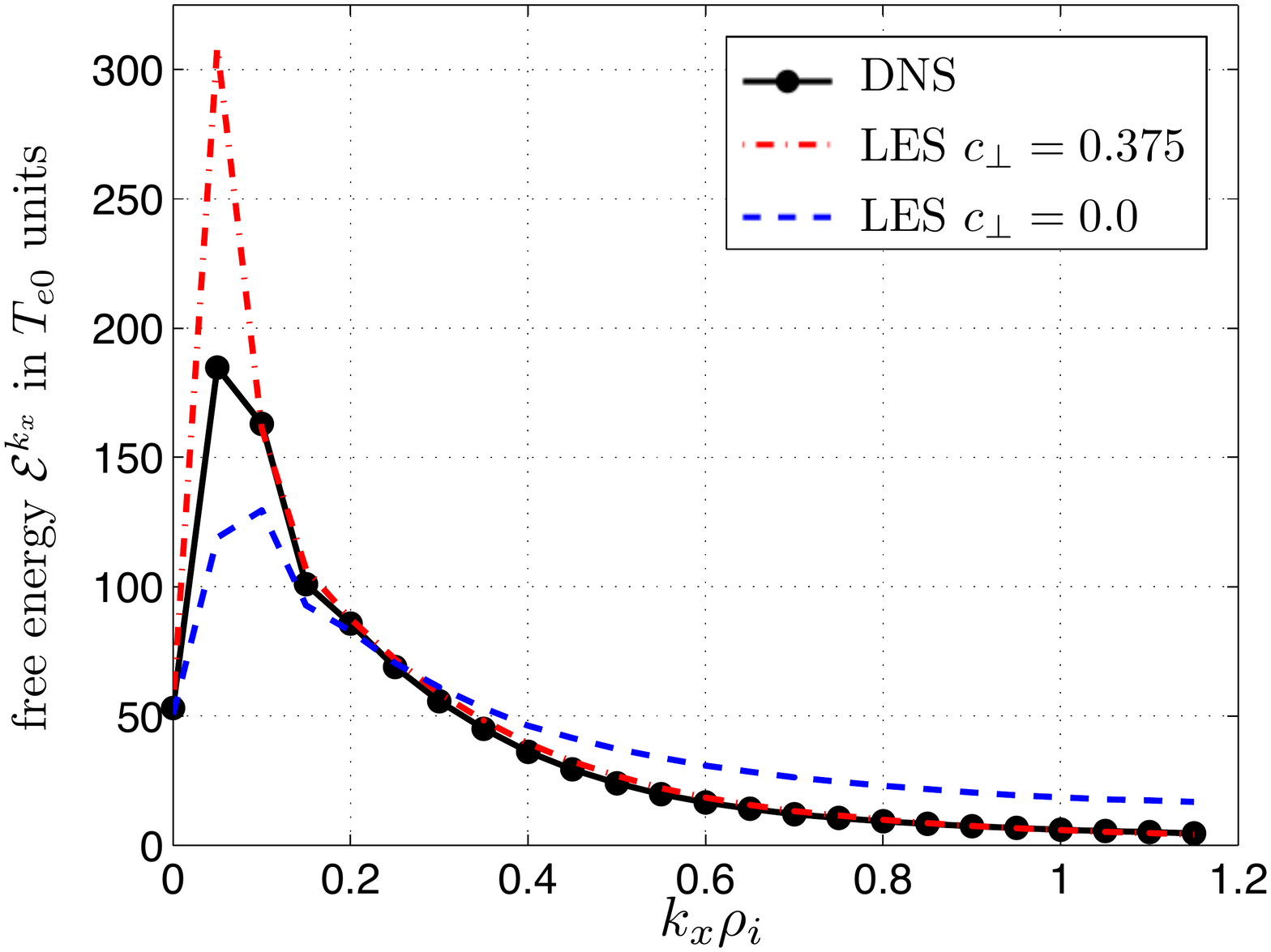}
\includegraphics[width=0.50\textwidth]{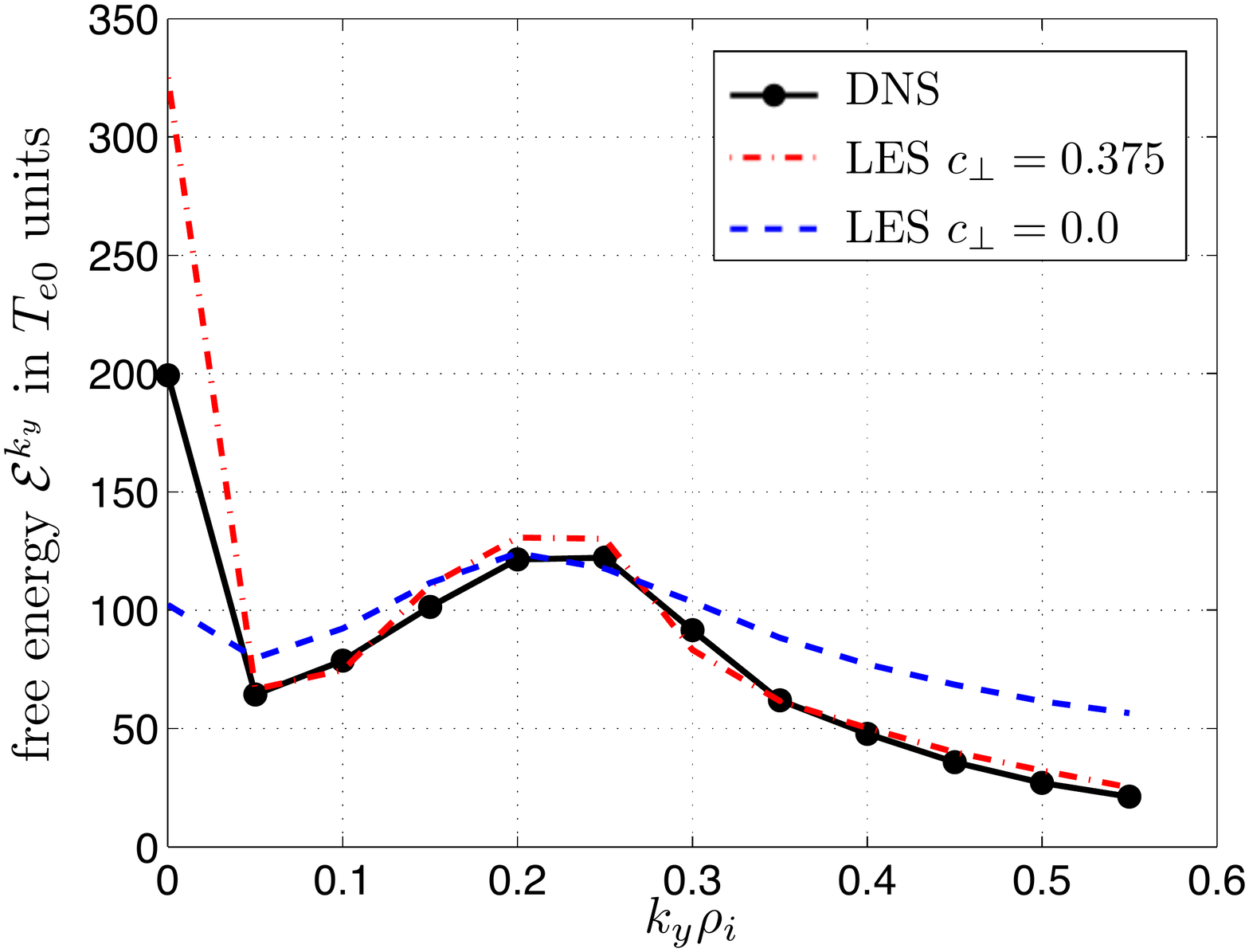}
\caption{\label{fig:cmp-LT60} Same as Figure~\ref{fig:cmp-LT80} with $\omega_{Ti} = 6.0$.}
\end{figure}

Such a test shows that the LES approach can reproduce qualitatively the expected phenomenology at a much lower cost than the DNS. However, it is important to assess the quantitative agreement between LES and DNS. For this reason, another comparison between LES and DNS has been performed for $\omega_{Ti}=8.0$. As shown in Fig.~\ref{fig:cmp-LT80}, the LES using the same model with $c_\perp = 0.375$ again reproduces the resolved free energy spectra obtained from the DNS reasonably well. In particular, there is a clear improvement when compared with the no-model run. The value of $\omega_{Ti}=8.0$ corresponds to a more turbulent state than $\omega_{Ti}=6.96$ and the LES appears to be fairly well robust in this regime. It should be acknowledged, however, that the situation is not fully satisfactory when turbulence strength is decreased. For instance, in Fig.~\ref{fig:cmp-LT60} ($\omega_{Ti} = 6.0$), the LES predictions, although still acceptable, starts to deviate significantly from the DNS results in the large scale range. This is reminiscent of a difficulty known in the development of model for LES in fluid turbulence. Very few models are capable to capture correctly the transition between laminar and turbulent flows. Probably, the very simple model proposed here also suffers from such a deficiency. 

\section{Discussion \label{discussion}}

The study presented here shows that the concept of LES can be extended to three-dimensional (in space) gyrokinetics. The very good agreement reported in Figs.~\ref{fig:scan-cperp} and \ref{fig:cmp-LT80} between fully resolved simulations and under-resolved simulations including a simple model for the filtered scales is encouraging. It shows that the model, calibrated for a given value of the temperature gradient in ITG turbulence, is able to reproduce the large scale spectra of the free energy for higher temperature gradient. It should be noted that such a test is quite demanding since the free energy has to be reproduced for each scale. Hence, the model has not only to be able to dissipate the correct amount of free energy, it also has to distribute the dissipation correctly amongst the different scales.

It should be acknowledged, however, that the robustness of the model has not been proved in the most general sense. For instance, when the turbulence level is lowered by decreasing the parameter $\omega_{Ti}$, the agreement between the under-resolved and the fully resolved simulations become less and less satisfactory -- although such behavior can be understood and even anticipated since, for low $\omega_{Ti}$, turbulence is not fully developed and the cascade picture starts to break down. An interesting extension of the present approach would then be to apply the dynamic procedure used to calibrate automatically the amplitude of sub-grid scale models in LES for fluid turbulence~\cite{germano_PoF_1991}. The dynamic procedure is known to be able to predict the transition between turbulence and laminar flows by automatically setting the model amplitude to zero in the laminar regime in which the small scales are not active. Hence, although the models are designed by using concepts valid for fully developed turbulence, the dynamic procedure seems to be able to extend their validity into totally different regimes. However, the implementation of the dynamic procedure in gyrokinetics is more intricate than the use of the simple model studied here. 

Finally, it is interesting to discuss the computational gain obtained in the LES simulation presented in the previous section. Although for the present case, the fully resolved reference simulation is not using a very large grid, the LES grid (and hence the required memory) can be reduced by 86\%. In terms of CPU time, the gain is even higher. Indeed, the simulation can be performed with a larger time step since the smallest scales are larger than in the reference simulation (in practice, with the grid resolution chosen in this study, the time step is increased by a factor of about two in the LES when compared to the DNS). As a consequence, the overall computational effort required for the LES runs appears to be more than 20 times smaller than in the DNS simulations. This finding indicates that the LES approach is quite promising in the context of gyrokinetics. Further studies along these lines, also exploring other types of plasma turbulence (driven, e.g., by trapped electron modes or electron temperature gradient modes), will be the focus of future work.

\acknowledgments{
The authors would like to thank G.~W.~Hammett, G.~G.~Plunk, T.~Tatsuno, and D.~R.~Hatch for very fruitful discussions. We gratefully acknowledge that the results in this paper have been achieved with the assistance of high performance computing resources on the HPC-FF system at J\"ulich, Germany. This work has been supported by the contract of association EURATOM - Belgian state.
}


\end{document}